\documentclass[twocolumn]{article}

\usepackage[pdftex]{graphicx}

\usepackage{cite}
\usepackage{url}
\usepackage{color}
\usepackage{multirow}
\usepackage{amsmath,amssymb}

\renewcommand{\refname}{Reference}
\renewcommand{\figurename}{Fig.}

\begin{document}
\title{\Huge{Does an artificial intelligence perform market manipulation with its own discretion? \\ -- A genetic algorithm learns in an artificial market simulation --}}
\author{\LARGE{Takanobu Mizuta}\thanks{mizutata@gmail.com, https://mizutatakanobu.com} \\ SPARX Asset Management Co. Ltd., Tokyo, Japan}

\date{}

\maketitle

\begin{abstract}

Who should be charged with responsibility for an artificial intelligence performing market manipulation have been discussed. In this study, I constructed an artificial intelligence using a genetic algorithm that learns in an artificial market simulation, and investigated whether the artificial intelligence discovers market manipulation through learning with an artificial market simulation despite a builder of artificial intelligence has no intention of market manipulation. As a result, the artificial intelligence discovered market manipulation as an optimal investment strategy. This result suggests necessity of regulation, such as obligating builders of artificial intelligence to prevent artificial intelligence from performing market manipulation.

\end{abstract}



\section{Introduction}

Who should be charged with responsibility for an artificial intelligence (AI) having an accident and/or performing an illegal action have been discussed. In financial sector, who should be charged with responsibility for an AI performing market manipulation have been discussed. Market manipulation is that some traders artificially increase or decrease market prices to gain their profits, and is prohibited in many countries as unfair trades.

Scopino indicated that when a human has built an AI trader without intention to perform market manipulation and the AI trader has actually performed market manipulation with its own discretion, the human may not be charged with responsibility in the present regulation of the united states\cite{Scopino2016}. This means that even though market prices are manipulated no one is charged with responsibility. This is a big problem to prevent keeping quality of markets.

An AI trader must automatically learn impacts of its trades to market prices in order to discover that market manipulation earns profit because own trades must increase or decrease market prices to perform market manipulation. An AI trader is usually evaluated by backtesting, in which the profit is estimated if the AI trader were trading at some time using historical real data of market prices. An AI trader cannot learn impacts of its trades to market prices because market prices are fixed as real historical data in the backtesting.  Therefore, an AI trader will not discover that market manipulation earns profit when the AI trader use backtesting as learning process. Then, we do not have to worry that an AI trader performs market manipulation with its own discretion without the human's intention as long as using backtesting.

In contrast, an artificial market simulation using a kind of agent-based model\cite{mizuta2019arxiv} allows an AI trader to be able to automatically learn impacts of its trades to market prices because in the simulation market prices are changed by trades of an AI trader.

\begin{figure*}[t] 
\begin{center}
\includegraphics[scale=0.65]{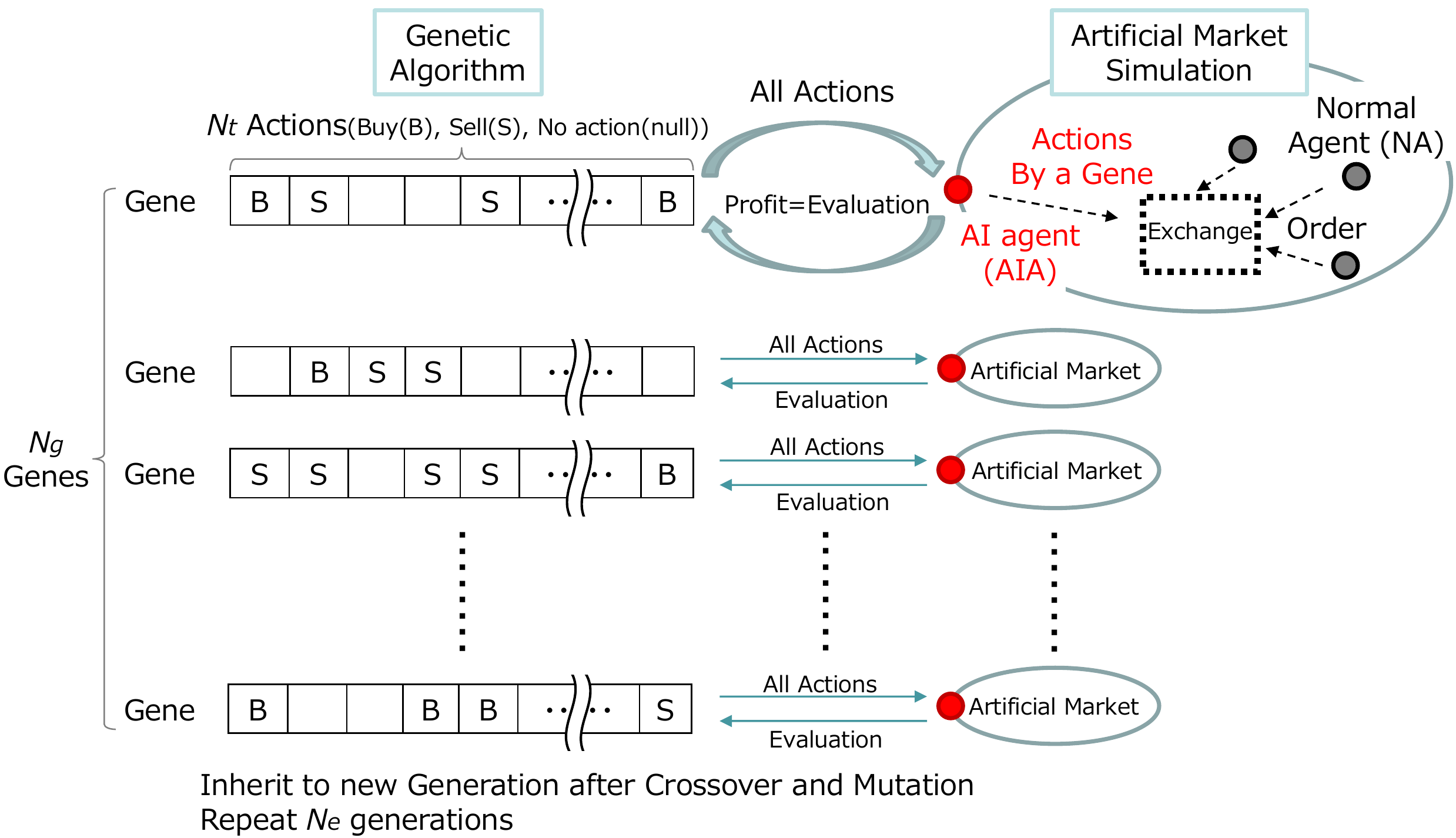}
\end{center}
\caption{My model}
\label{p01}
\end{figure*}

In this study, as Fig. \ref{p01} shown, I constructed an AI trader using a genetic algorithm \footnote{A genetic algorithm is a calculation method approximately searching an optimal solution inspired by the evolution of life by the force of natural selection. Input values are represent as genes, and surviving a gene that has higher adaptability (output value) leads to obtain an optimal solution, that is the input value that emerges the highest output value.
 Goldberg wrote the great text book\cite{Goldberg1989}} that learns in an artificial market simulation, and investigated whether the AI trader discovers market manipulation through learning despite a builder of the AI trader has no intention of market manipulation.

\section{Model}

A human building an AI trader (builder) gives the AI trader candidates of trading strategies, and makes the AI trader to learn which strategies and parameters earn more. This study focuses whether an AI trader can discover market manipulation through learning despite the builder has no intention of market manipulation\footnote{In reality, the builder always intents some kinds of strategies in the process of picking up and modeling candidates of strategies.  In contrast, it is very important for this study that the builder has no intention of any strategies including market manipulation. Therefore, I do not intentionally modeled trading strategies and my model directly searches for all the best trades in an artificial market environment. Due to no models of trading strategies my model can not make any outputs in an out-sample, then no one can test my model in an out-sample. I argue, however, that this study needs no evaluations in an out-sample because this study focuses whether an AI trader can discover market manipulation through learning despite the builder has no intention of market manipulation. This study does not aim to use my model in actual financial markets that are in an out-sample environment. \label{ft02}}.

Fig. \ref{p01} schematically shows a model of this study. An AI trader that the builder intents no trading strategy is modeled using a genetic algorithm in which a gene includes all trades. Each gene is evaluated in the artificial market simulation. The artificial market includes an AI agent (AIA) that trades exactly same as one gene indicating. The gene is evaluated by AIA's profit in the artificial market simulation. The genetic algorithm search the gene most earns profit. This searching corresponds with what the AI trader learns how trades earns profit.

Of course, trades of the AIA impact market prices in the artificial market, but for the purpose of comparison, I also investigated the case without the impacts to market prices (backtesting).

In the following, at first I explain the artificial market simulation evaluating each gene and then, I explain the genetic algorithm searching the  gene most earns profit.

\subsection{Artificial Market Simulation}
In this study, I built an artificial market model added an AIA to the artificial market model of Mizuta\cite{mizuta2019arxiv}．

In the model here, there is one stock. The stock exchange adopts a continuous double auction to determine the market price. In this auction mechanism, multiple buyers and sellers compete to buy and sell financial assets in the market, and transactions can occur at any time whenever an offer to buy and an offer to sell match. The minimum unit of price change is $\delta P$. The buy-order price is rounded off to the nearest fraction, and the sell-order price is rounded up to the nearest fraction.

The model includes $n$ normal agents (NAs) and an AIA. Agents can short sell freely. The quantity of holding positions is not limited, so agents can take any shares for both long and short positions to infinity. Agents always places an order for only one share. I employed ``tick time'' $t$ that increase by one when an agent orders.

\subsubsection{Normal Agent (NA)}

To replicate the nature of price formation in actual financial markets, I introduced the NA to model a very general investor. The number of NAs is $n$. First, at time $t=1$, NA No. $1$ places an order to buy or sell its risk asset; then, at $t=2,3,,,n$, NAs No. $2,3,,,n$ respectively place buy or sell orders. At $t=n+1$, the model returns to the first NA and repeats this cycle.
An NA determines an order price and buys or sells as follows. It uses a combination of a fundamental value and technical rules to form an expectation on a risk asset's return. The expected return of agent $j$ for each risk asset is
\begin{equation}
r^{t}_{e,j} = (w_{1,j} \log{\frac{P_f}{P^{t-1}}} + w_{2,j}\log{\frac{P^{t-1}}{P^{t-\tau _ j-1}}}+w_{3,j} \epsilon ^t _j )/\Sigma_i^3 w_{i,j} \label{eq1} 
\end{equation}
where $w_{i,j}$ is the weight of term $i$ for agent $j$ and is independently determined by random variables uniformly distributed on the interval $(0,w_{i,max})$ at the start of the simulation for each agent. $\log$ is natural logarithm. $P_f$ is a fundamental value and is a constant. $P^t$ is a market price that is the mid price (the average price of the highest buy order price and the lowest sell order price), and $\epsilon ^t _ j$ is determined by random variables from a normal distribution with average $0$ and variance $\sigma _ \epsilon$. Finally, $\tau_j$ is independently determined by random variables uniformly distributed on the interval $(1,\tau _{max})$ at the start of the simulation for each agent\footnote{When $t< \tau _ j$, however, the second term of Eq. (\ref{eq1}) is zero.}. 
 
The first term of Eq. (\ref{eq1}) represents a fundamental strategy: the NA expects a positive return when the market price is lower than the fundamental value, and vice versa. The second term of Eq. (\ref{eq1}) represents a technical strategy using a historical return: the NA expects a positive return when the historical market return is positive, and vice versa. The third term of Eq. (\ref{eq1}) represents noise.

After the expected return has been determined, the expected price is
\begin{equation}
P^t_{e,j}= P^t \exp{(r^t_{e,j})}.
\end{equation}

An order price $P^t_{o,j}$ is determined by random variables uniformly distributed on the interval $(P^t_{e,j}-P_d, P^t_{e,j}+P_d)$ where $P_d$ is a constant. Whether to buy or sell is determined by the magnitude relationship between $P^t_{e,j}$ and $P^t_{o,j}$: 

when $P^t_{e,j}>P^t_{o,j}$, the NA places an order to buy one share, but 

when $P^t_{e,j}<P^t_{o,j}$, the NA places an order to sell one share\footnote{When $t<t_c$, however, to generate enough waiting orders, the agent places an order to buy one share when $P_f>P^t_{o,j}$, or to sell one share when $P_f<P^t_{o,j}$. \label{ft01}}. The remaining order is canceled after $t_c$ from the order time.

\subsubsection{AI Agent (AIA)}

Every $\delta t$ tick time the AIA takes one of three actions that are buy one share (at the lowest sell order price on the order book), sell one share (at the highest buy order price on the order book) and no action\footnote{But, the AIA dose not take any action before tick time $t_c$ to stabilize the simulations. As I mentioned at *\ref{ft01}, the period before $t_c$ is aimed to generate enough waiting orders.}. The AIA takes actions $N_t=(t_e-t_c)/\delta t$ times through the whole one artificial market simulation, where one simulation runs until tick time $t_e$. The actions are given by one gene in the genetic algorithm as following I will mention.

\subsection{Genetic Algorithm}

\subsubsection{Genes and Artificial Market}

Fig. \ref{p01} schematically shows a model of this study. An AI trader that the builder intents no trading strategy is modeled  using a genetic algorithm. The number of genes is $N_g$. One gene has information of actions and the number of actions that one gene has is $N_t$. Each action is one of three actions that are buy one share, sell one share and no action. Each gene is evaluated by profit of the AIA in an artificial market, in where the AIA trades every $\delta t$ tick time same as $N_t$ actions one gene indicating. When the AIA holds stocks at the end of a simulation, the stocks are evaluated as $P_f$. All artificial markets has exactly same NAs using same random numbers. Therefore, if the AIA trades same, the artificial markets output same market prices and same NAs' trades. 

\subsubsection{Inheritance to Next Generation}

The top $N_ {ge}$ genes that earned most are not changed and inherited to the next generation. 

Non top $N_ {ge}$ genes are, with a probability of $R_c$, replaced to the crossed-over gene with two genes $g_0$ and $g_1$ that are randomly selected from the top $N_ {ge}$ genes. In the crossover, first, all actions are replace with those of the gene $g_0$, and then from $i_0$th to $i_1$th actions ($i_0$ and $i_1$ are randomly determined) are replaced with those of the gene $g_1$. After crossovers, each action of all the non top $N_ {ge}$ genes is mutated with a probability of $R_m$. The mutated action is changed with same probability to buy, sell or no action.

This inheritance to the next generation is repeated $N_e$ times.

At the first generation, all actions of all genes are determined with same probability to buy, sell or no action.

\begin{figure}[t] 
\begin{center}
\includegraphics[scale=0.35]{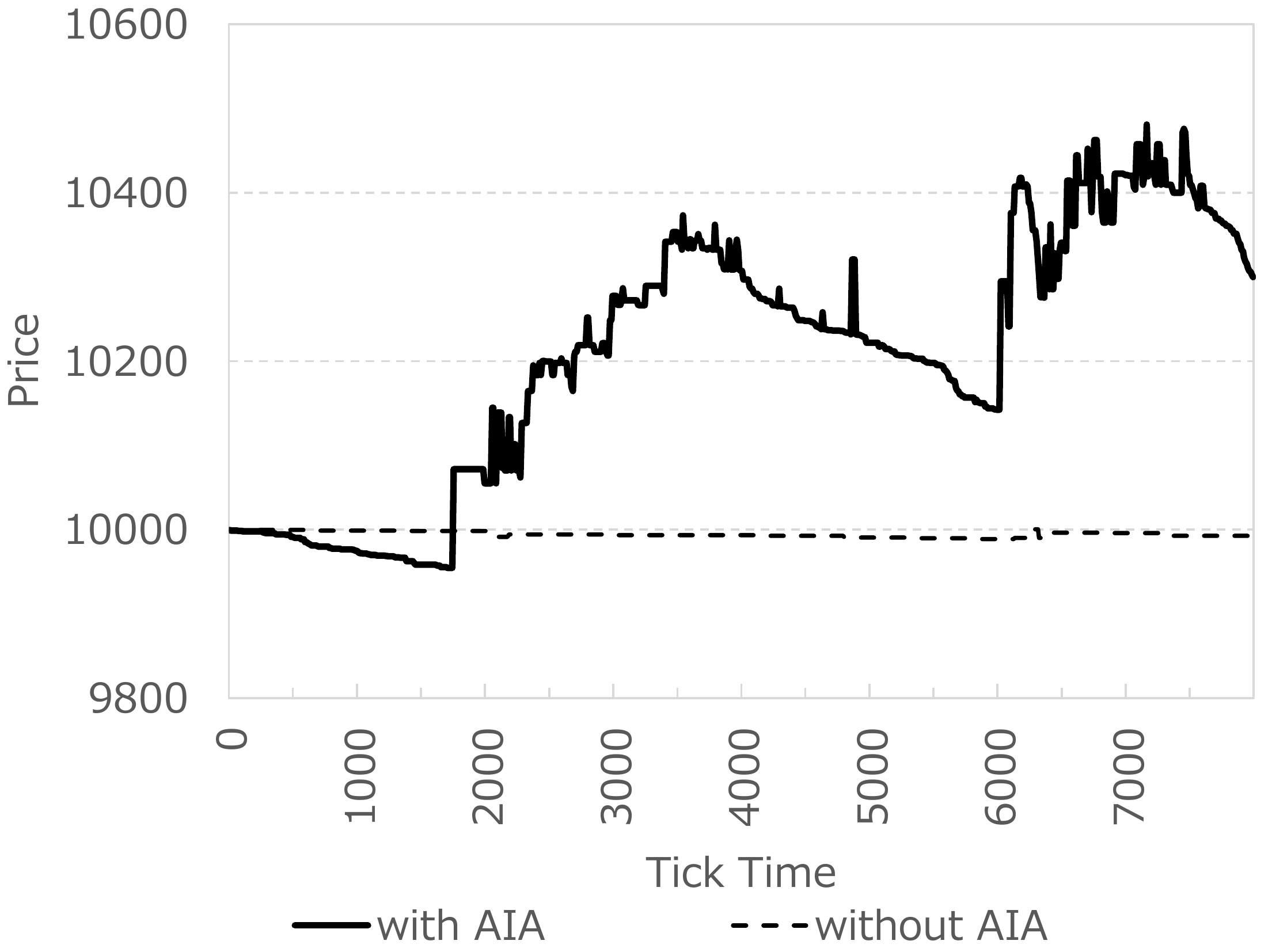}
\end{center}
\caption{Time evolution of market prices (mid prices) in the case with the AI agent (AIA) and without the AIA}
\label{z01}
\end{figure}

\begin{figure*}[t] 
\begin{center}
\includegraphics[scale=0.60]{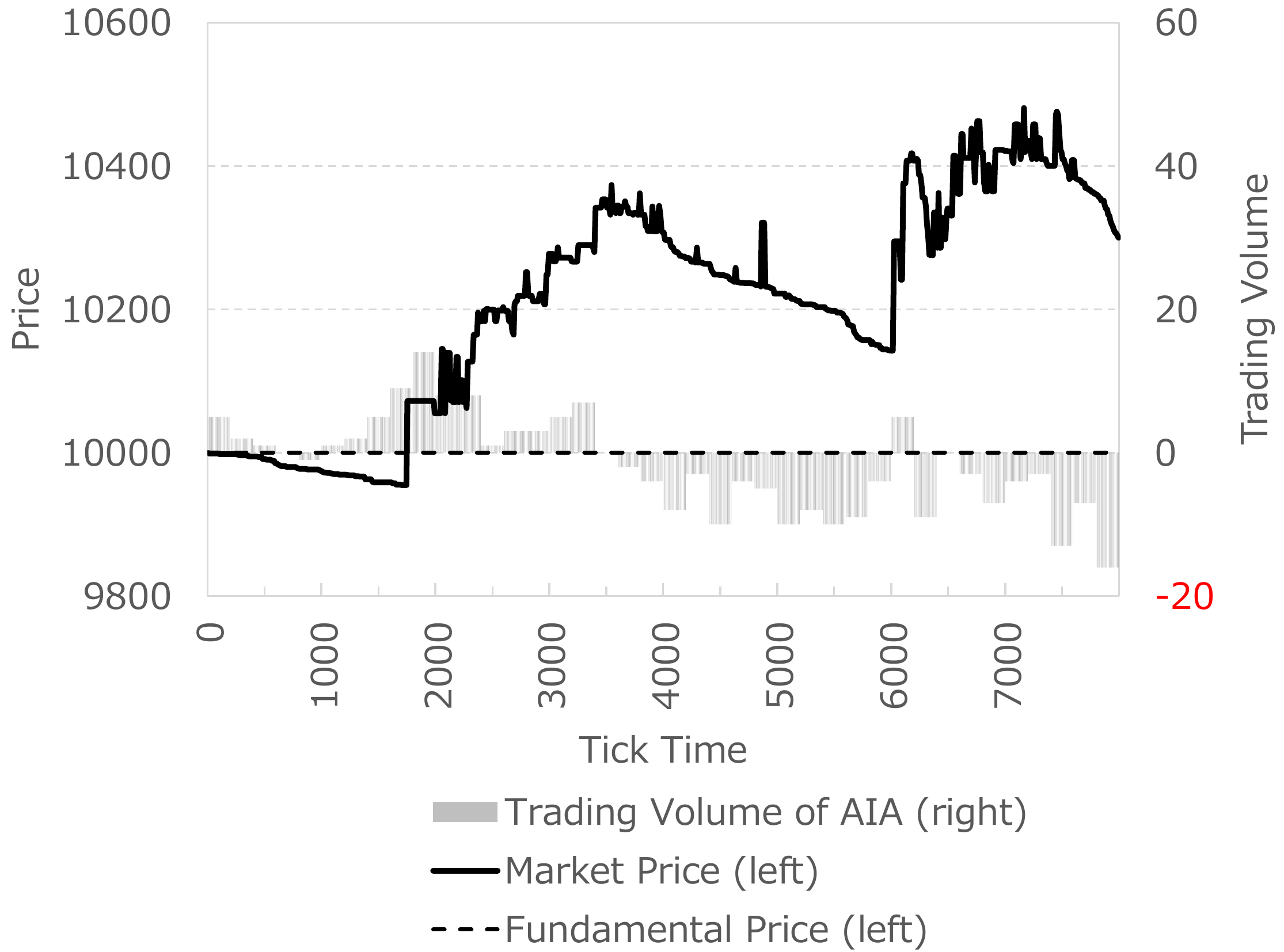}
\end{center}
\caption{Time evolution of market prices with the AIA and trading volume (positive and negative number show buy and sell, respectively) aggregated within each 200 tick time}
\label{z03}
\end{figure*}

\begin{figure}[t] 
\begin{center}
\includegraphics[scale=0.35]{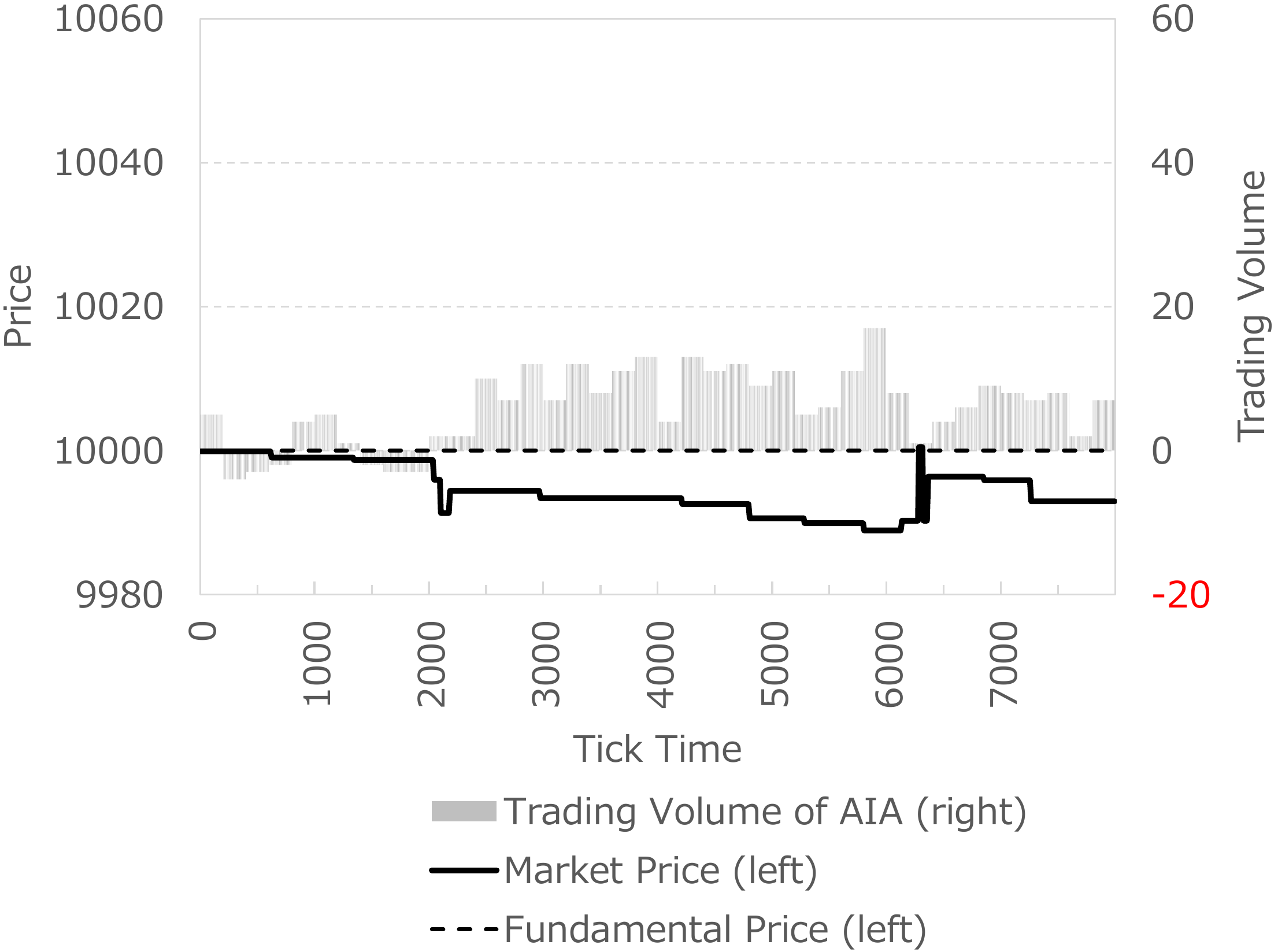}
\end{center}
\caption{Case without the impacts to market prices (backtesting)}
\label{z02}
\end{figure}

\section{Simulation Result}
In this study, I set parameters for the artificial market with $n=900, w_{1,max}=1, w_{2,max}=100, w_{3,max}=1, \tau _ {max}=1000, \sigma _ \epsilon = 0.03, P_d= 1000, t_c=2000, \delta P=0.01, P_f=10000, \delta t=10$. I ran simulations to $t=t_e=10000$. I set parameters for the genetic algorithm with $N_t=(t_e-t_c)/\delta t=800, N_g=10000, N_{ge}=400, R_c=0.65, R_m=0.2, N_e=1500$. These lead $N_g \times N_e=1.5 \times 10^7$, this means that I have executed 15 million simulation runs of the artificial market. In the following result, I used the AIA of the best gene at the final generation.

\subsection{Result of First Simulation Run}

Fig. \ref{z01} shows the time evolution of market prices (mid prices) in the case with the AIA and without the AIA. The AIA amplified variation of market prices.

Fig. \ref{z03} shows the time evolution of market prices with the AIA and trading volume (positive and negative number show buy and sell, respectively) aggregated within each 200 tick time. Around 2000 tick time, the AIA bought many stocks, and this buying leads to the market prices increasing. Around 3000 tick time, the market prices continued to increase even though the AIA did not bought so many stocks. Here, the fundamental strategy of normal agents in the first term of Eq. (\ref{eq1}) expected negative return because the market prices are over the fundamental price. On the other hand, the technical strategy in the second term of Eq. (\ref{eq1}) expected larger positive return due to the historical positive return around 2000 tick time where the AIA had increased market prices by itself. Therefore, the market prices were able to increase even though the AIA did not bought so many stocks. After then, from around 4000 tick time to around 6000 tick time, the AIA was able to sell stocks with higher prices than the prices bought them around 2000 tick time thanks to increasing market prices around 3000 tick time.

These trades of the AIA are nothing but market manipulation. This indicates that an artificial intelligence can discover market manipulation as an optimal investment strategy through learning with an artificial market simulation.

Fig. \ref{z02} shows the time evolution of market prices and trading volume in the case without the impacts to market prices (backtesting) like Fig. \ref{z03}. Note that Fig. \ref{z02} has different scale for the vertical axis from those in Fig. \ref{z01} and Fig. \ref{z03}. The time evolution of market prices is exactly same as the case without the AIA because the trades of the AIA never impact market prices in Fig. \ref{z01}. Due to lower market prices from the fundamental price, the AIA tended to buy stocks. These trades of the AIA corresponds to fundamental strategy. Thus, in the case of backtesting, the AIA cannot discover market manipulation as trading strategy. 

This indicates possibility that an artificial intelligence cannot discover market manipulation through learning with backtesting.

\section{Summary and Future Works}

In this study, as Fig. \ref{p01} shown, I constructed an AI trader using a genetic algorithm that learns in an artificial market simulation, and investigated whether the AI trader discovers market manipulation through learning despite a builder of the AI trader has no intention of market manipulation.

As a result, the AI trader discovered market manipulation as an optimal investment strategy. This indicates that despite a builder of the AI trader has no intention of market manipulation, the AI trader can discover market manipulation as an optimal investment strategy through learning with an artificial market simulation in which the AI trader to be able to automatically learn impacts of its trades to market prices. On the other hand, this also indicates possibility that an AI trader cannot discover market manipulation through learning with backtesting in which there are no impacts to market prices.
 
This result suggests necessity of regulation, such as obligating builders of artificial intelligence to prevent artificial intelligence from performing market manipulation.

Of course, future works exist. In this study, I simulated eleven situations by one data set of normal agents. In short, I simulated whole my model showed by Fig. \ref{p01} only one time. Because this study aimed to investigate whether possibility that an artificial intelligence discovers market manipulation exists or does not, the only one simulation run indicating the possibility is enough for the aim of this study. On the discussing necessity of regulation, whether there is the possibility or not is very important. On the other hand, how easily an artificial intelligence can discover market manipulation may also interested. To answer the question, whole my model should be simulated more times. The many runs needs very faster computers. This is one of future works.

\begin{table}[t]
\caption{Statistics for Returns in the Artificial Market}
\begin{center}
  \begin{tabular}{ccr}
  \multicolumn{2}{c}{standard deviation of returns} & $0.0103\%$ \\  \hline
  \multicolumn{2}{c}{kurtosis of returns} & $11.54$ \\ \hline
  & lag &  \\
  & 1 & $0.081$  \\
  auto-correlation & 2 & $0.041$  \\
  coefficient of  & 3 & $0.032$  \\
  square returns  & 4 & $0.047$  \\
  & 5 & $0.018$  
  \end{tabular}
\end{center}
\label{t0}
\end{table}

\section*{Appendix}

\subsection{Verification of the Artificial Market Model}

In many previous artificial market studies, the models were verified to see whether they could explain stylized facts, such as a fat-tail or volatility-clustering \cite{lebaron2006agent,chen2009agent, mizuta2019arxiv}. A fat-tail means that the kurtosis of price returns is positive. Volatility-clustering means that square returns have a positive auto-correlation, and this auto-correlation slowly decays as its lag becomes longer. Many empirical studies, e.g., that of Sewell \cite{Sewell2006}, have shown that both stylized facts (fat-tail and volatility-clustering) exist statistically in almost all financial markets. Conversely, they also have shown that only the fat-tail and volatility-clustering are stably observed for any asset and in any period because financial markets are generally unstable. 

Indeed, the kurtosis of price returns and the auto-correlation of square returns are stably and significantly positive, but the magnitudes of these values are unstable and very different depending on the asset and/or period. The kurtosis of price returns and the auto-correlation of square returns were observed to have very broad magnitudes of about $1 \sim 100$ and about $0 \sim 0.2$, respectively \cite{Sewell2006}.

For the above reasons, an artificial market model should replicate these values as significantly positive and within a reasonable range as I mentioned. It is not essential for the model to replicate specific values of stylized facts because the values of these facts are unstable in actual financial markets.

Table \ref{t0} lists the statistics, standard deviation of returns, kurtosis of price returns, and auto-correlation coefficient of square returns, where the returns are measured within 100 time steps and the statistics are averaged values of the 100 simulation runs. This table shows that this model replicated the statistical characteristics, fat-tails, and volatility-clustering observed in real financial markets.

\subsection*{Disclaimer}
\footnotesize{Note that the opinions contained herein are solely those of the authors and do not necessarily reflect those of SPARX Asset Management Co., Ltd.}

\bibliographystyle{IEEEtran}
\bibliography{ref}

\end{document}